\documentclass[aps,twocolumn,showpacs,floatfix]{revtex4}
\usepackage{graphicx}
\begin{document}

\title {Rotating Bose-Einstein condensates with attractive interactions}
\author{G. M. Kavoulakis,$^1$ A. D. Jackson,$^2$ and Gordon Baym$^{3,4}$}
\affiliation{$^1$Mathematical Physics, Lund Institute of Technology, 
                 P. O. Box 118, S-22100 Lund, Sweden \\
             $^2$Niels Bohr Institute, Blegdamsvej 17, DK-2100 Copenhagen
		 \O, Denmark \\
             $^3$Department of Physics, University of Illinois at
        Urbana-Champaign, 1110 West Green Street, Urbana, Illinois 61801 \\
             $^4$NORDITA, Blegdamsvej 17, DK-2100 Copenhagen \O, Denmark}
\date{\today}

\begin{abstract}

    We examine the phase diagram of a Bose-Einstein condensate of atoms,
interacting with an attractive pseudopotential, in a quadratic-plus-quartic
potential trap rotating at a given rate. Investigating the behavior of the
gas as a function of interaction strength and rotational frequency of the
trap, we find that the phase diagram has three distinct phases, one with vortex
excitation, one with center of mass excitation, and an unstable phase in which
the gas collapses.

\end{abstract}
\maketitle

\section{Introduction}

    The velocity field of a superfluid is irrotational.  As a result, such
fluids develop vortex states when rotated, and the circulation of the
superfluid velocity around any closed path is quantized \cite{PS}.  Trapped
Bose-Einstein condensates of alkali-metal atoms provide ideal systems for
testing these ideas.  Remarkably, they permit experimentalists not only to
vary the form of the trapping potential but also to change the sign of the
scattering length associated with the effective atom-atom interaction.

    In a harmonic trap, vortices in Bose-Einstein condensates with repulsive
interactions are always singly-quantized \cite{Rokhsar,KMP}.  If, however, the
effective interaction is attractive or the trapping potential has a different
shape \cite{Dal}, the situation can change dramatically.  The recent
experiment of Ref.\,\cite{Dal} investigated the behavior under rotation of an
effectively repulsive Bose-Einstein condensate confined in an anharmonic trap.
As shown in the theoretical studies of
Refs.\,\cite{Lundh,Fetter,tku,KB,AA,JK,FB,JKL}, vortices can be
multiply quantized in any trap that grows faster than quadratically.  Further,
Refs.\,\cite{WGS,BM,PP} have demonstrated that the angular momentum is carried
by center-of-mass motion in an effectively-attractive condensate that rotates
in a harmonic trap.

    In this study we investigate the phase diagram of an
anharmonically-trapped Bose gas in the case of an effective attractive atomic
interaction as a function of the rotation frequency of the trap and of the
strength of the dimensionless coupling constant, $Na/Z$, where $N$ is the
total number of atoms, $a$ is the s-wave scattering length, and $Z$ is height
of the gas along the rotation axis.  Both the sign and the magnitude of the
scattering length can be adjusted using Feshbach resonances.  The magnitude of
the coupling constant can also be controlled by changing either the number of
atoms in the trap or the trap frequencies.  For sufficiently strong attractive
interactions, such systems are unstable against collapse \cite{BP,UL,EM}.  
However, as long as the interaction is balanced by the zero-point motion of 
the atoms in the trap (and also by the kinetic energy associated with the 
vortices in the case of a rotating gas), the gas can exist in a metastable 
state.

    In discussing the effects of rotation on the metastability of a gas with
attractive interactions we must distinguish two physical situations, whether
the trap is being rotated at a given angular frequency -- the principal case
we consider here -- or whether the gas is set into rotation, and then the
rotation of the trap is switched off.  The former case, the analog of the
Meissner effect in superconductors, examines the equilibrium states of the
system in the presence of rotation.  The latter examines the metastability of
supercurrents in the absence of a driving force.  As stressed by Leggett
\cite{Leggett}, superfluid flow in a system with effectively attractive
interactions is not stable in the absence of external rotation.  Thus the
rotating states that we find here to be metastable in a rotating trap should
lose angular momentum and cease rotating once the rotation of the trap is
switched off.  The details of how the angular momentum associated with
superfluid flow in such systems is lost is beyond the scope of this paper.
Examples of the breakup of a vortex are given in Refs.~\cite{U1,U2}.  Angular
momentum can also be continuously taken out of the flow through interaction of
the flow with normal fluid.

    Anharmonic traps allow a rich structure of three phases depending on the
strength of the attractive interactions and on the rotational frequency.  The
first is an unstable phase in which the gas collapses to a more dense state.
The second phase is characterized by the angular momentum being carried by the
motion of the center of mass.  The third phase involves either a mixed state
of multiply and singly quantized vortices or a pure state of multiple
quantization.  The general structure of the corresponding phase diagram is
shown schematically in Fig.\,1.  References \cite{EL} have investigated this
problem using both variational and exact numerical techniques.  It is crucial
that the trap be anharmonic to have such a structure.  In a harmonic trap,
repulsive interactions ensure that the critical frequency for rotation is
smaller than the trap frequency and thus permit the formation of vortices
\cite{Rokhsar}.  Attractive interactions increase the critical frequency for
rotation to a value larger than the trap frequency with the result that the
centifugal force cannot be balanced by the restoring force of the trap and the
atoms fly apart.
\begin{figure}
\begin{center}
\includegraphics[width=8cm,height=6cm]{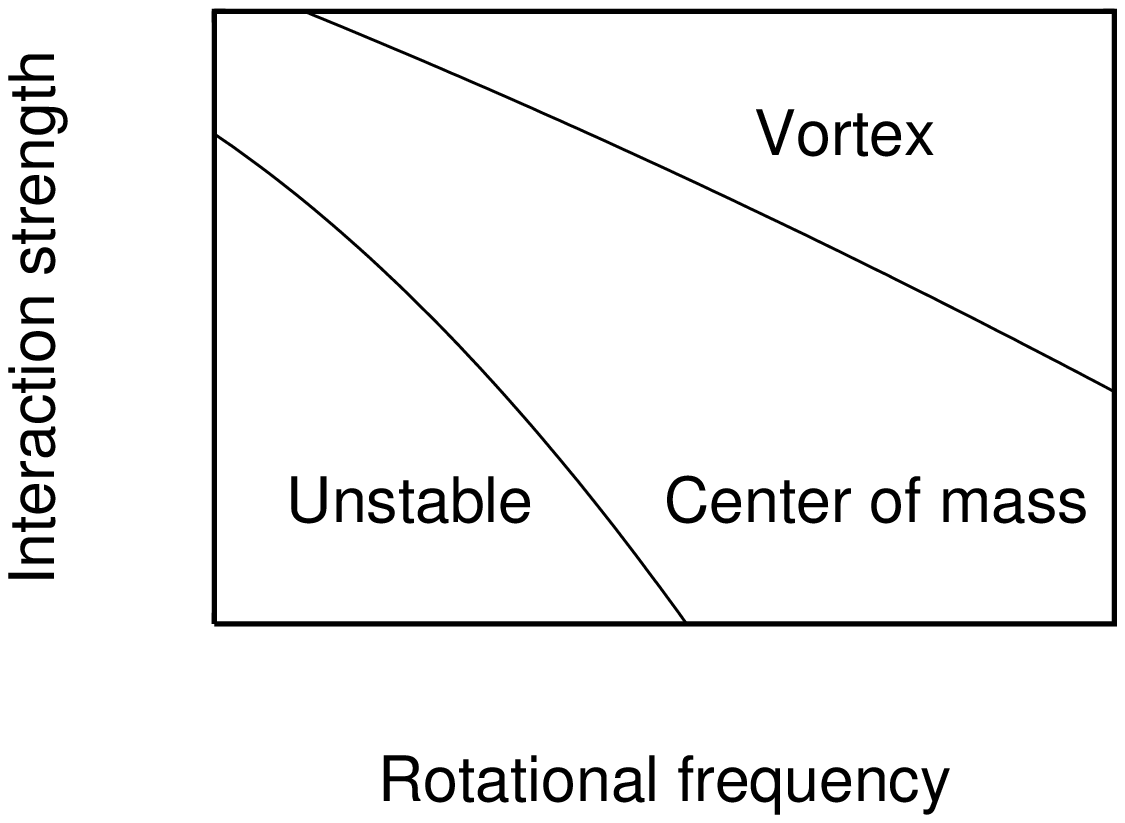}
\begin{caption}
{Schematic phase diagram of a rotating Bose-Einstein condensate trapped in
an anharmonic potential.  The origin is the upper left corner.  The vertical
axis is the strength of the attractive effective atom-atom coupling constant
and the horizontal axis is the rotational frequency of the trap.}
\end{caption}
\end{center}
\label{FIG1}
\end{figure}

    In analyzing the problem we consider atoms interacting via a short-range
effective interaction,
\begin{eqnarray}
   V_{\rm int} =
      \frac 1 2 U_{0} \sum_{i \neq j} \delta({\bf r}_{i} - {\bf r}_{j}).
\label{v}
\end{eqnarray}
Here, $U_0 = 4 \pi \hbar^2 a/M$ is the strength of the effective two-body
interaction, where $M$ is the atomic mass.  We consider an anharmonic trapping
potential of the form
\begin{equation}
   V(\rho,z) = \frac 1 2 M \omega^2 \rho^2
     \left[1 + \lambda \left(\frac {\rho} {d_0}\right)^2 \right] + V_z(z).
\label{anh}
\end{equation}
Here, $\omega$ is the trap frequency of the harmonic potential
perpendicular to the $z$ axis (which is assumed to be the axis of
rotation) and $d_0 = (\hbar/M \omega)^{1/2}$ is the oscillator length.
The dimensionless constant $\lambda$ is small (e.g., $\lambda \approx 5
\times 10^{-3}$ in the experiment of Ref.~\cite{Dal}) and $V_z$ is
the trapping potential along the $z$ axis.  Denoting the excitation energy
of the first excited state along the $z$ axis by $\Delta E_z$, we assume that
\begin{equation}
  n U_0 \ll \hbar \omega \ll \Delta E_z,
\label{ineq}
\end{equation}
where $n$ is the typical atom density.  The inequality between the left
and the right terms implies that the cloud is in its lowest state of motion
along the $z$ axis, and the problem thus becomes effectively two dimensional.
Furthermore, under the above conditions, the typical density is $\sim N/d_0^2
Z$.  The interaction energy $n U_0$ is $\sim \sigma a \hbar \omega$, where
$\sigma = N/Z$ is the atom density per unit length.  The assumption of small
$nU_0$ in the left inequality of Eq.\,(\ref{ineq}), is equivalent to
$\sigma a \ll 1$.

    In the following section, we examine the phase where the rotating gas
forms vortices.  In Sec.\,III we study the phase in which the angular momentum
is carried by the center of mass, and in Sec.\,IV examine the unstable phase.
Section V discusses the general features of the phase diagram, and Sec.\,VI
summarizes our conclusions.

\section{Vortex phase}

    For sufficiently weak interactions, the behavior of the system under
rotation is dominated by the anharmonicity of the trap, and the many-body wave
function is a product of single-particle states describing multiply -- or
singly-quantized vortex states.  Assuming weak interaction ($\sigma a \ll 1$)
and weak anharmonicity ($\lambda \ll 1$), we can restrict our attention to a
basis of eigenstates of the harmonic potential with zero radial excitations
and angular momentum $m \hbar$.  (We now set $\hbar = M = \omega = 1$ for
convenience.  We denote many-body states by capital letters and
single-particle states by small letters.)
The basis states are
\begin{equation}
    \psi_m({\rho, \phi}) =
    \frac 1 {\sqrt{\pi m!}} {\tilde z}^m e^{-|{\tilde z}|^2/2},
\label{phim}
\end{equation}
where $m$ is a non-negative integer, and ${\tilde z} = \rho e^{i \phi}$.
The symmetrized mean-field many-body state with total angular momentum $L$ can
then be expanded in this basis as
\begin{equation}
  \Psi_{\rm v}({\bf r}_1,\dots,{\bf r}_N) =
    \prod_{i=1}^N \sum_{m=0}^{\infty}
    \frac {c_{m} {\tilde z}_i^{m}} {\sqrt{m!}} \Psi_0,
\label{mbsv}
\end{equation}
with $\sum_{m=0}^{\infty} m |c_m|^2= L/N$, and $\sum_{m=0}^{\infty} |c_m|^2=1$,
where $\Psi_0$ is the many-body state of the nonrotating cloud,
\begin{equation}
      \Psi_0 = \frac 1 {\pi^{N/2}}
      \exp \left( \sum_{i=1}^N -|{\tilde z}_i|^2/2 \right)
      \Phi_z(z_1, \dots, z_N).
\label{mbgs}
\end{equation}
Here, $\Phi_z$ is the ground-state many-body wavefunction of the cloud
along the $z$ axis.  Henceforth, we neglect $\Phi_z$, since it plays
no role in our analysis.

    The energy of the system in the state $\Psi_{\rm v}$ was calculated in
Refs.\,\cite{JK,JKL}.  Here, we note only certain features related to the
trapping potential.  If ${\bf w}_i = {\bf r}_i - {\bf R}$ are relative
coordinates (with ${\bf R}$ the center of mass coordinate), the potential
energy can be written as
\begin{eqnarray}
   V({\rho}_1, \dots, {\rho}_N) = \frac12 \left[
        \sum_{i=1}^N \left( {w_i}^2 + \lambda w_i^4 \right) \phantom{XXXXXX}
      \right. \nonumber \\ \left.
          + N R^2 + N \lambda R^4
   + 4 \lambda R^2 \sum_{i=1}^N w_i^2 \right].
\label{anhmb}
\end{eqnarray}
In the state $\Psi_{\rm v}$, the first terms, involving $w_i$, are of
order $N\hbar\omega$ and $\lambda N\hbar\omega$; by contrast the last three
terms in Eq.\,(\ref{anhmb}) are of order $\hbar \omega$ (i.e., unity),
$\lambda \hbar \omega/N$, and $\lambda \hbar \omega$, respectively, since
$N\langle R^2\rangle$ is of order unity.  Therefore, in the frame rotating
with angular frequency, $\Omega$, the energy per particle of a
multiply-quantized vortex state,
\begin{equation}
  \Psi_{\rm v}({\bf r}_1,\dots,{\bf r}_N) =
      \prod_{i=1}^N
          \frac {{\tilde z}_i^{m}} {\sqrt{m!}} \Psi_0,
\label{mbsv2}
\end{equation}
is
\begin{eqnarray}
  \frac {E'_{\rm v}} {N \hbar \omega} &=& 1+ m \left(1 - \frac {\Omega}
  {\omega}\right)
        + \frac \lambda 2 (m+1) (m+2) + \nonumber \\ &&+
                        \sigma a \frac {(2 m)!} {2^{2m} (m!)^2},
\label{totexprot}
\end{eqnarray}
as in Refs.\,\cite{JK,JKL}.

    As shown in these references, if $\sigma |a|$ exceeds some critical value,
the state containing a multiply quantized vortex, described by $\psi_m$,
is unstable against becoming a state of mixed angular
momentum of the form
\begin{equation}
   \psi = c_{m-1} \psi_{m-1} + c_{m} \psi_{m} + c_{m+1} \psi_{m+1}.
\label{mst}
\end{equation}
The dashed curves in Fig.~2 show this phase boundary with $\lambda =
0.05$, for $m = 1$ to $9$ from left to right.  In each of the regions above
the dashed lines, the energy of the gas in the rotating frame is minimized for
$|c_m| = 1$, and $ |c_{m\pm 1}| = 0$.
\begin{figure}
\begin{center}
\includegraphics[width=8cm,height=5cm]{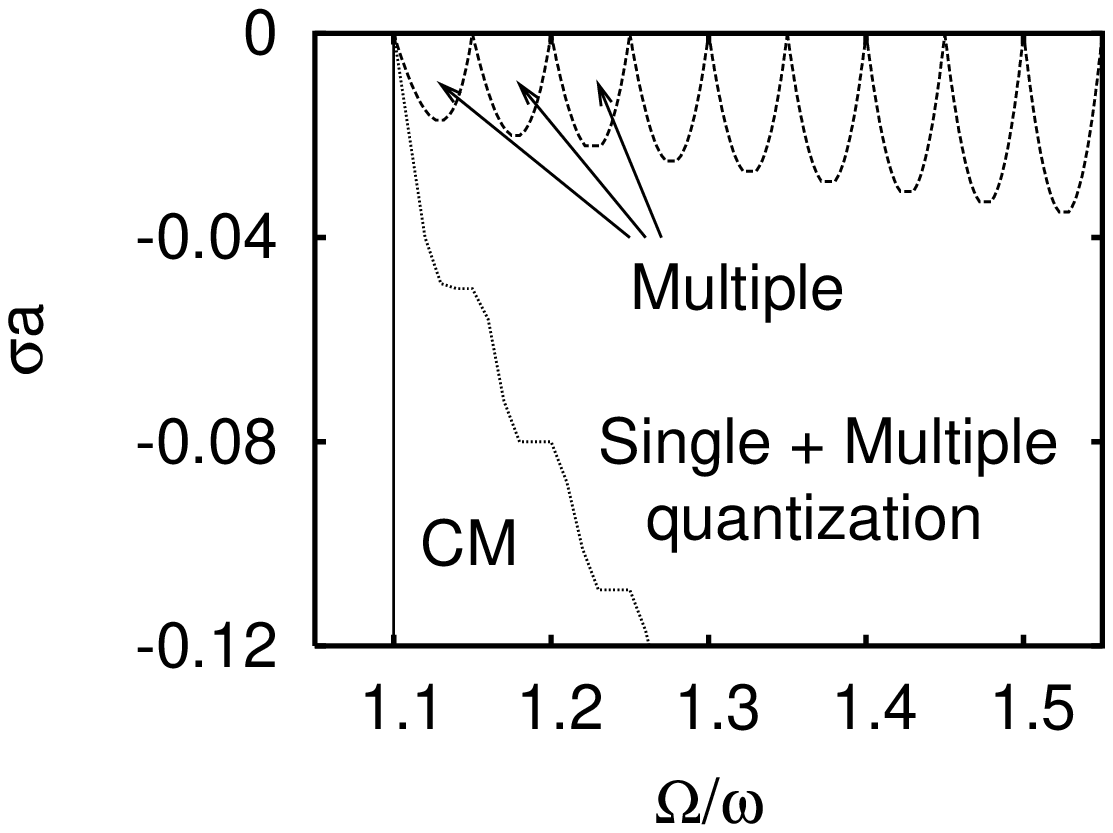}
\begin{caption}
{The phase diagram of a Bose-Einstein condensate trapped in a
quadratic-plus-quartic potential in the $\Omega/\omega$ -- $\sigma a$ plane
for $\lambda = 0.05$.  The vertical line denotes the critical frequency for
exciting the center of mass (CM), and the lower dotted line gives the boundary
between vortex and center-of-mass motion.  To the left of the vertical
line, the system does not rotate.  The higher dashed line denotes the
boundary between pure multiply-quantized vortex states with winding numbers $m
= 1$ to $9$ and a mixed phase consisting of multiply and singly quantized
vortices.}
\end{caption}
\end{center}
\label{FIG2}
\end{figure}

\begin{figure}
\begin{center}
\includegraphics[width=7cm,height=5cm]{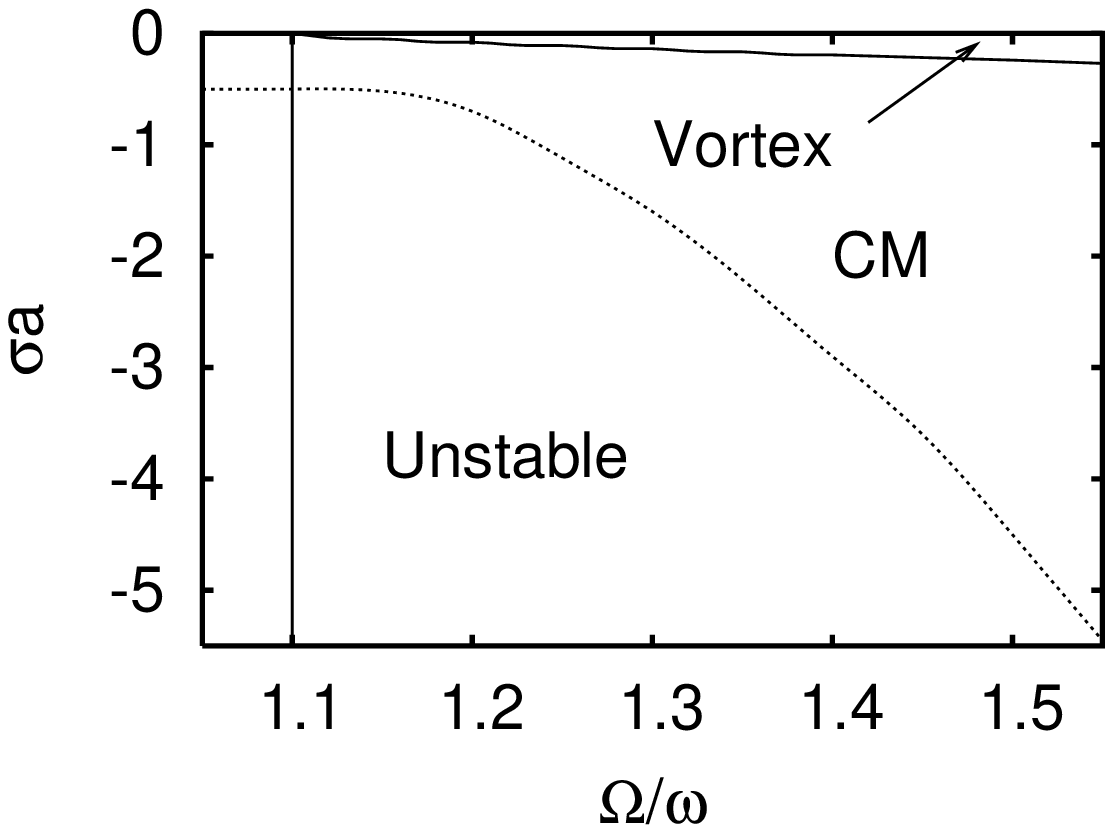}
\begin{caption}
{The same graph as in Fig.\,2 for a wider range of values of $\sigma a$.
The higher curve separating the vortex phase and the center of mass phase
is the lower curve shown in Fig.\,2.}
\end{caption}
\end{center}
\label{FIG3}
\end{figure}

\section{Center of mass excitation}

    If the scattering length $a$ is sufficiently negative, we expect the
angular momentum to be carried by the center of mass motion \cite{WGS,BM}.
The many-body wavefunction is then,
\begin{equation}
  \Psi_{\rm cm}({\bf r}_1,\dots,{\bf r}_N) =
     \frac 1 {\sqrt{N^L L!}} \left( \sum_{i=1}^N z_i \right)^L \Psi_0.
\label{mbscom}
\end{equation}
where $L$ is non-negative.  Calculating the energy per particle of the
system in the state $\Psi_{\rm cm}$ in the rotating frame, we find that
\begin{equation}
    \frac {E'_{\rm cm}} {N \hbar \omega} = 1 + l \left(1 - \frac {\Omega}
   {\omega}\right) + \frac {\lambda} 2 (l^2+4l+2) + \sigma a,
\label{totexprot2}
\end{equation}
with $l=L/N$. The term $\lambda l^2/2$ comes from the quartic component
$\propto R^4$ in Eq.\,(\ref{anhmb}), the term $2\lambda l$ results from the
component $\lambda \sum_i w_i^4$, while the term $\lambda$ results
from $4 \lambda R^2 \sum_i w_i^2$.  The final term in
Eq.\,(\ref{totexprot2}) is the interaction energy of a non-rotating cloud.
Since center of mass motion does not affect the relative coordinates,
$\langle V_{\rm int}\rangle$ is unaltered.

    Differentiating Eq.\,(\ref{totexprot2}) with respect to $l$ we find that
the critical frequency of rotation of the cloud (for center of mass
excitation) is \cite{EL},
\begin{equation}
      \Omega_c / \omega = 1 + 2 \lambda,
\label{omcritical}
\end{equation}
which is shown, for $\lambda = 0.05$, as the vertical curve in Figs.~2 and 3.

  For sufficiently negative interaction strength the energy of $\Psi_{\rm cm}$
becomes smaller than that of $\Psi_{\rm v}$, and we expect a phase
transition from the vortex phase investigated in Sec.\,II to the center of
mass state.  The overlap between these two states is exponential small in the
number of atoms.  In the limit of a large number of atoms, this transition is
discontinuous in contrast to the transitions examined in Sec.\,II.  In the
case of a pure giant vortex given by Eq.\,(\ref{mbsv2}), for example, the
overlap is
\begin{equation}
  \langle \Psi_{\rm cm} | \Psi_{\rm v} \rangle = \left(\frac {L!} {N^L
    (m!)^N}\right)^{1/2}
\label{overlap1}
\end{equation}
with $L = m N$. In the limit $N \to \infty$,
\begin{eqnarray}
  \langle \Psi_{\rm cm} | \Psi_{\rm v} \rangle
  &=& (2 \pi m)^{1/2} \left( \frac {m^m} {e^m m!} \right)^{N/2}
\nonumber \\
  &=& (2 \pi m )^{-N/4},
\label{overlap1aa}
\end{eqnarray}
where last equality applies for $m \gg 1$.

    The phase boundary in the $\Omega/\omega$ -- $\sigma a$ plane is 
approximately the line along which $E'_{\rm v} = E'_{\rm cm}$, given by
Eqs.~(\ref{totexprot}) and (\ref{totexprot2}). This result is shown as 
the lower (higher) curve in Fig.\,2 (Fig.\,3).  In fact, the true phase 
boundary lies even lower since the energy for the single-particle mixed 
states of the form $\psi = \cdots c_{m-1} \psi_{m-1} + c_{m} \psi_{m} + 
c_{m+1} \psi_{m+1} \cdots$ is necessarily smaller than that for $\psi = 
\psi_m$ in this region.

\section{Unstable phase}

    Sufficiently large attractive interactions render the cloud unable to
support itself, and a dense atomic state results.  In a nonrotating system,
this instability occurs when $\sigma |a|$ is on the order of unity
\cite{BP,UL}.  As we show below, in a rotating cloud this value is even more
negative because of the kinetic energy of the rotational motion \cite{ben}.

To calculate the phase boundary, we start with the energy in the
rotating frame in the state $\Psi_{\rm cm}$, Eq.\,(\ref{totexprot2}), but
regard the corresponding oscillator length as a variational parameter.
We previously assumed that the oscillator length is
fixed and given by $(\hbar/N M \omega)^{1/2}$.  Here, we perform the same
calculation assuming that the oscillator length is $\beta
(\hbar/N M \omega)^{1/2}$, where $\beta$ is real and positive. The result is
\begin{equation}
  \frac {E'_{\rm cm}} {N \hbar \omega} = \frac {1 + l} 2 (\beta^2 + \frac 1
   {\beta^2}) - l \frac {\Omega} {\omega}
            + \frac {\lambda} 2 [l^2 \beta^4+ (4l+2)\beta^2] + \frac {\sigma
    a}  {\beta^2}.
\label{totexprot2var}
\end{equation}
For a non-rotating cloud, the value of $\beta$ that minimizes the energy is
\begin{equation}
  \beta_0 = \left( \frac {1 + 2 \sigma a} {1 + 2 \lambda} \right)^{1/4},
\end{equation}
which implies that the critical value for collapse is $\sigma a = -1/2$.

    Differentiating $E'_{\rm cm}$ with respect to $l$ we obtain
\begin{equation}
  l = \left[ \frac {\Omega} {\omega} -
  \frac 1 2 \left(\beta^2 + \frac 1 {\beta^2} \right)
    - 2 \lambda \beta^2 \right] \frac 1 {\lambda \beta^4},
\end{equation}
so that the critical frequency for center of mass excitation is
\begin{equation}
  \frac {\Omega_c} {\omega} = \frac 1 2 \left(\beta^2 + \frac 1 {\beta^2}
  \right) +   2 \lambda \beta^2.
\end{equation}

    This expression reduces to Eq.\,(\ref{omcritical}) when $\beta\to 1$ in
the limit of weak interactions, $\sigma a \ll 1$ and $\lambda \ll 1$.  The
point where $E'_{\rm cm}$, expressed in terms of $\beta$, $\Omega/\omega$,
$\sigma a$, and $\lambda$, ceases to have a local minimum as function of
$\beta$, signals the collapse of the cloud.  The result for $\lambda = 0.05$
and varying $\Omega/\omega$ and $\sigma a$ is shown in the lower line of
Fig.~3.
\begin{figure}
\begin{center}
\includegraphics[width=7cm,height=5cm]{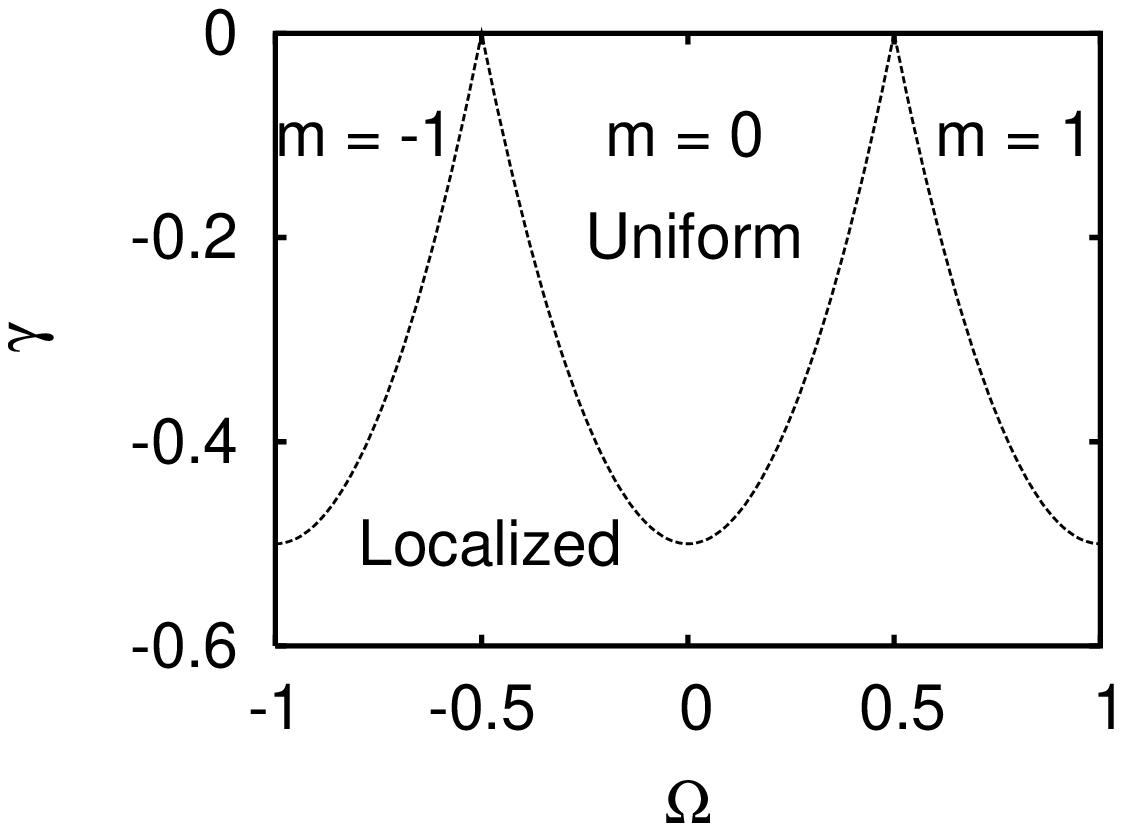}
\begin{caption}
{The phase diagram of an effectively attractive Bose-Einstein condensate
confined in a toroidal trap, from Ref.\,\cite{G}. Here $\gamma$ is the
ratio between the interaction energy and the kinetic energy, corresponding
to $\sigma a$ in the present problem, and $\Omega$ is the dimensionless
rotational frequency of the torus. The function $\gamma(\Omega)$ is periodic.}
\end{caption}
\end{center}
\label{FIG4}
\end{figure}

\begin{figure}
\begin{center}
\includegraphics[width=3cm,height=3cm]{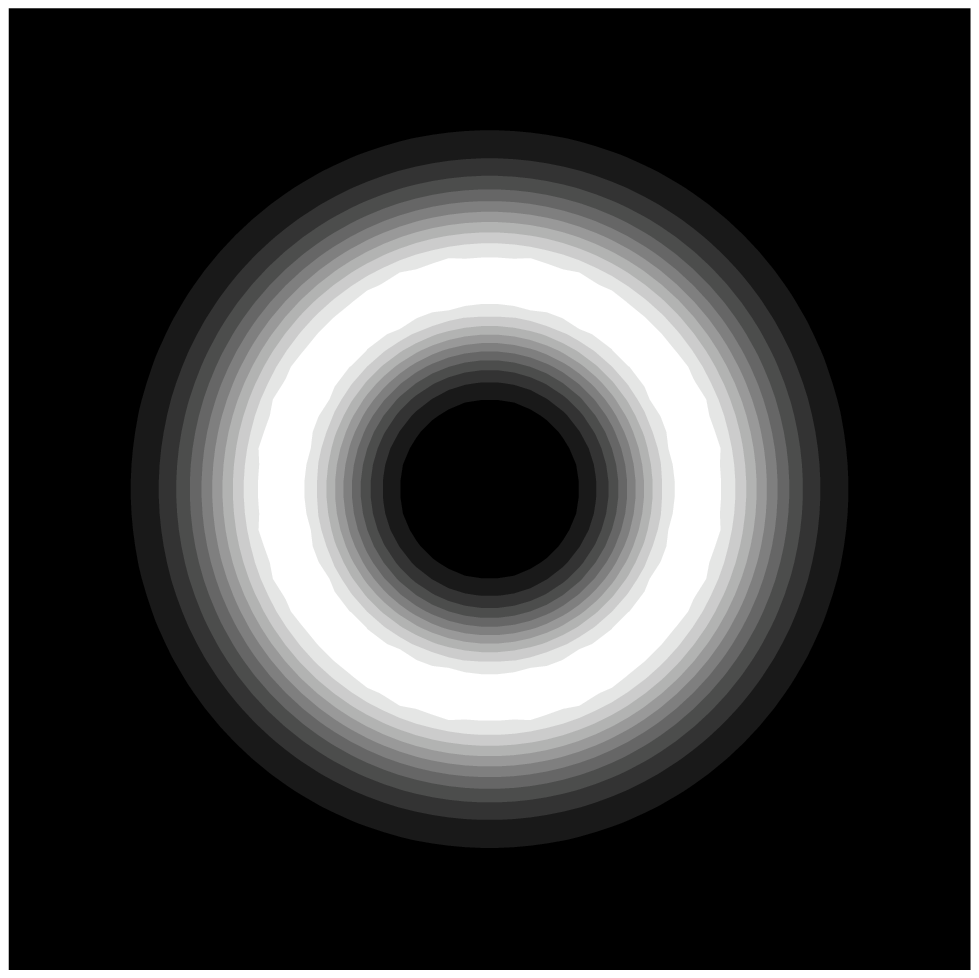}
\includegraphics[width=3cm,height=3cm]{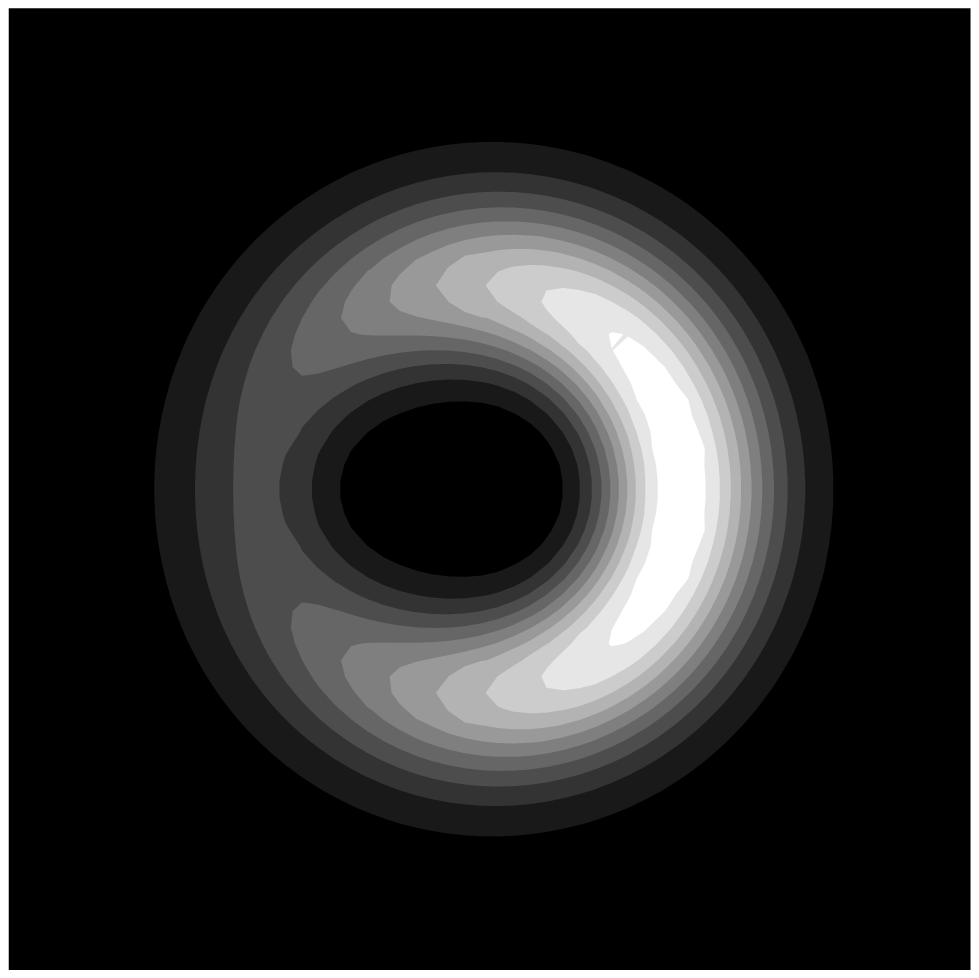}
\includegraphics[width=3cm,height=3cm]{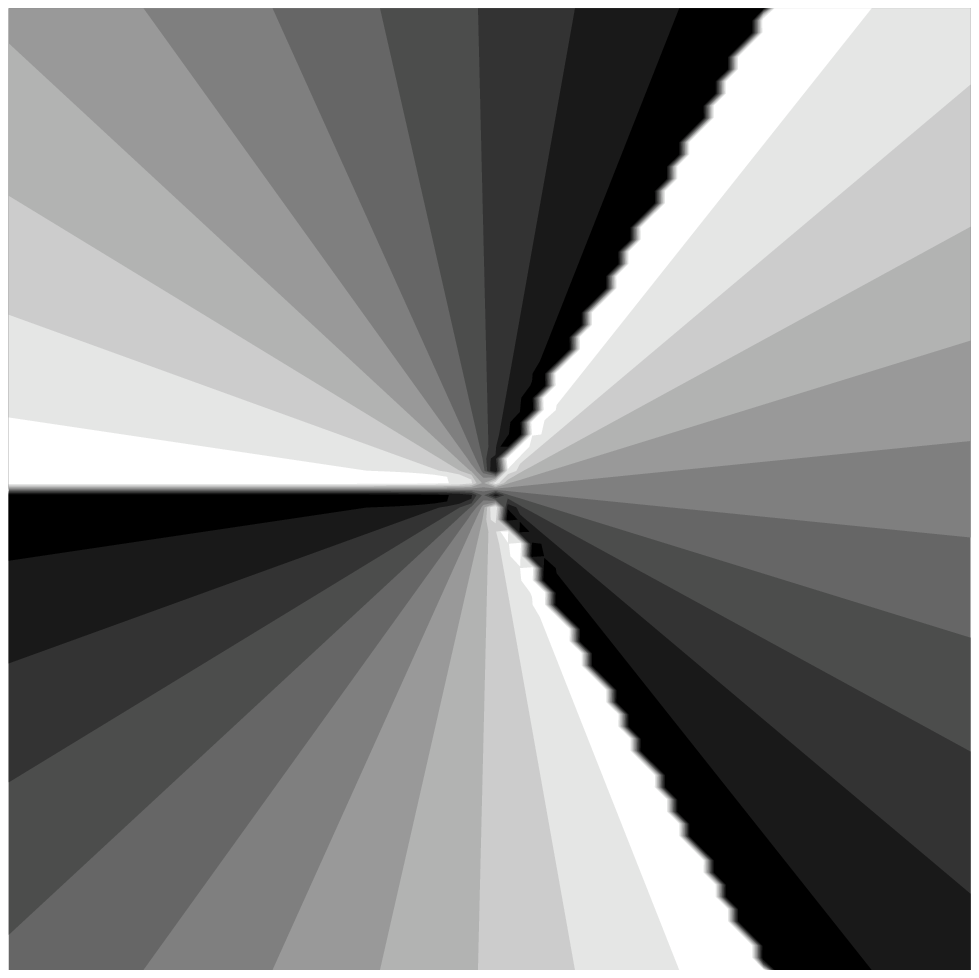}
\includegraphics[width=3cm,height=3cm]{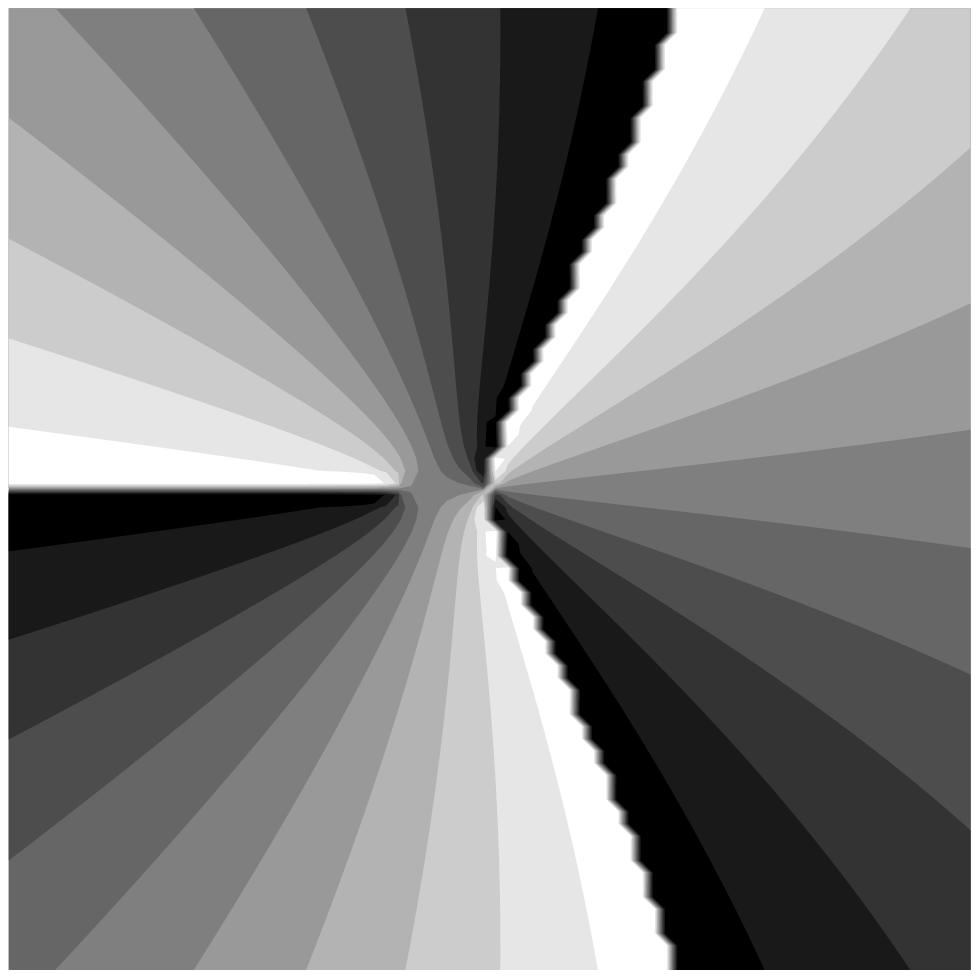}
\begin{caption}
{Equipotential plots of the density (higher) and phase (lower) of the
condensate in the state of the form of Eq.\,(\ref{mst}) for $m = 3$.  In the
left graphs $c_2 = c_4 = 0$, while in the right ones all $c_m$ are nonzero.
The left graphs correspond to the uniform phase of the one-dimensional
problem, while the right ones correspond to the localized phase.  The axes
extend from $-4 d_0$ up to $4 d_0$.}
\end{caption}
\end{center}
\label{FIG5}
\end{figure}

\section{Features of the phase boundary}

    The calculated phase diagram of Fig.\,2 is universal \cite{JK,JKL} in the
sense that the same graph would be obtained for $\lambda'=\alpha \lambda$ by
rescaling the axes by $\alpha$, i.e.,
$(1-\Omega/\omega)\to\alpha(1-\Omega/\omega)$ and $\sigma a \to \alpha \sigma
a$.  Further, the dashed line indicating the phase boundary between pure
multiple quantization and the mixed phase is exact in the limit of small
$\lambda$ and $\sigma |a|$ \cite{JK,JKL}.  In the phases of multiple
quantization, the energy in the rotating frame is minimized (i.e., the energy
is an extremum and its second derivative with respect to admixtures of other
states is always positive).  Thus, even though the effective interaction is
attractive in this system, the vortices of the driven system are stable
against small perturbations.  Furthermore, as one decreases $\Omega/\omega$
crossing the segments of fixed $m$, there is a continuous path along which the
energy decreases towards the absolute minimum of the energy (in the lab
frame).  This implies that there can be no persistent currents, in agreement
with the toy model presented in Sec.\,VI of Ref.\,\cite{Leggett}.

    The higher phase boundary in Fig.~2 closely resembles that of a rotating
Bose-Einstein condensate confined in a one-dimensional toroidal trap, with an
effective attractive interaction between the atoms.  This system was
investigated in Refs.\,\cite{U2,G}; its phase diagram is shown in Fig.~4.  In
this one-dimensional problem the instability of states of the form $\phi_m =
e^{i m \phi}/\sqrt{2 \pi}$ is actually towards the combination $c_{m-1}
\phi_{m-1} +c_{m} \phi_{m} + c_{m+1} \phi_{m+1}$, which is the one-dimensional
analogue of our problem.  [See Eq.\,(\ref{mst}).] The reason for this
resemblance is the form of the effective potential felt by the atoms,
\begin{eqnarray}
  V_{\rm eff} = V - M \Omega^2 \rho^2/2,
\end{eqnarray}
which has a mexican-hat shape for $\Omega > \omega$.  We identify the
phase of pure multiple quantization as the uniform state of the
one-dimensional problem and the other two phases as the localized state of the
one-dimensional case.  References \cite{EL} exclude the possibility of a state
of the form of Eq.\,(\ref{mst}).  However, at least for the parameter range
considered here close to the phase boundary, there is also a mixed phase for
which the order parameter has the form of Eq.\,(\ref{mst}).

    When $c_{m-1} = c_{m+1} = 0$, these states describe multiply-quantized
vortices of winding number $m$, and the density is homogeneous, as shown on
the left upper graph of Fig.~5 (for $m=3$.)  We see in the lower left graph a
triply quantized vortex at the origin.  On the other hand, when $c_{m-1}$ and
$c_{m+1}$ are nonzero, these states describe a combination of a
multiply-quantized vortex ($m=2$) at the center of the cloud with winding
number $m-1$, here 2 in the lower right figure, one singly-quantized vortex
just to left close to the origin, and another not shown, much further from the
center, where the density is very low.  As shown in the upper right graph of
Fig.\,5 (for $m=3$), the density is now inhomogeneous (and resembles that
corresponding to the center-of-mass phase).  This is an energetically
favourable configuration because the effective atom-atom attraction prefers
the inhomogeneity around the minimum of the effective potential $V_{\rm eff}$.
The reason why the minima in the phase boundary of Fig.\,2 decrease with
increasing $\Omega/\omega$ (as opposed to the one-dimensional case) is that
the radius of the toroidal trap corresponding to $V_{\rm eff}$ increases with
$\Omega/\omega$.

    We also note that with increasing $\Omega$, at fixed $\sigma a$, center of
mass excitation always occurs before quantized vortices are formed.  The
states $\Psi_{\rm cm}$ and $\Psi_{\rm v}$ cannot be connected by low-order
application of any single-particle operator because of their small overlap, as
noted above.  Thus, even in the region where $\Psi_{\rm v}$ has a lower energy
than $\Psi_{\rm cm}$, the system can remain in $\Psi_{\rm cm}$ for times which
can be very long (i.e., that scale at least linearly with the number of atoms
$N$.)  The calculation of this time scale remains a difficult open question.
On the other hand, if one first rotates the gas above the condensation
temperature and then cools down, the system will end up in the phases shown in
Fig.~2.

\section{Summary}

    We have calculated the phase diagram for a rotating Bose-Einstein
condensate in a quadratic-plus-quartic potential when the effective
interaction between the atoms is effectively attractive.  For very weak
interactions there is a phase of (pure) multiply quantized vortices.  In this
phase the energy of the gas in the rotating frame is minimized and the vortex
states are stable against weak perturbations.  For somewhat more attractive
interactions, there is a mixed phase of single and multiple quantization.  For
even more attractive interactions, the system carries its angular momentum via
center of mass excitation.  Finally for even stronger attraction, the cloud
cannot support itself and collapses to a dense atomic state.

\vskip1pc

    Author GMK wishes to thank Emil Lundh and Ben Mottelson for useful
discussions.

\end{document}